\documentclass[12pt]{article}
\usepackage[mathscr]{eucal}
\usepackage{epsfig,amsfonts}
\usepackage[fleqn]{amsmath}
\usepackage{amsthm,amssymb}
\usepackage{graphicx}
\usepackage{hhline}
\usepackage{cite}

\makeatletter
\@addtoreset{equation}{section}
\makeatother

\def\be{\begin{equation}}
\def\ee{\end{equation}}
\def\bdm{\begin{displaymath}}
\def\edm{\end{displaymath}}
\def\bea{\begin{eqnarray}}
\def\eea{\end{eqnarray}}

\def\sgn{{\rm sgn}}

\def\ri{{\rm i}}

\def\XXint#1#2#3{{\setbox0=\hbox{$#1{#2#3}{\int}$}
    \vcenter{\hbox{$#2#3$}}\kern-.5\wd0}}

\newcommand{\rd}{\mbox{d}}
\newcommand{\re}{\mbox{e}}

\begin{document}

\begin{titlepage}
\begin{flushright}
RUNHETC-2012-10\\
\end{flushright}

\begin{center}
\begin{LARGE}

{\bf The integrable  harmonic map problem versus     Ricci flow}

\end{LARGE}
\vspace{1.3cm}

\begin{large}

{\bf  Sergei  L. Lukyanov}$^{1,2}$

\end{large}

\vspace{1.cm}

${}^{1}$NHETC, Department of Physics and Astronomy\\
     Rutgers University\\
     Piscataway, NJ 08855-0849, USA\\
\vspace{.2cm}
and\\
\vspace{.2cm}
${}^{2}$L.D. Landau Institute for Theoretical Physics\\
  Chernogolovka, 142432, Russia\\
\vspace{1.0cm}

\end{center}

\vspace{2cm}
\begin{center}
\centerline{\bf Abstract} \vspace{.8cm}
\parbox{15.5cm}
{
We construct a zero-curvature representation
for  a   four-parameter family of   non-linear sigma models with  a Kalb-Ramond term.
The one-loop renormalization  is performed  that
gives rise to a new set of   ancient and eternal solutions to the
Ricci flow with torsion.  Our analysis provides an explicit  illustration
of the  role of the dilaton field 
for the renormalization 
of the non-linear sigma model. 
}

\end{center}

\vfill

\end{titlepage}
\newpage

\section{Introduction}

Let ${\cal M}_D$  be  a
$D$-manifold (the {\it target space})   equipped  with a Riemannian metric  $G$  and
an affine connection.
Consider the system of 
PDE which describes 
a map  of the two-dimensional  worldsheet $\Sigma=(x^0,x^1)$
into the affine-metric manifold
\bea\label{sooisa}
\partial_+\partial_-X^{\mu}+{  \Gamma^{\mu}}_{\nu\sigma}\  \partial_+X^\nu\partial_-X^\sigma=0 \ .
\eea
Here  it is assumed
that $\Sigma$
is equipped with  a Minkowski metric, 
$\partial_\pm=\frac{1}{2}\,
(\partial_0\pm\partial_1)$ and 
${  \Gamma^{\mu}}_{\nu\sigma}$ stands for  Christoffel symbol of the  connection.
For a  general target space background, Eqs.\eqref{sooisa}
cannot be derived from the variational principal. 
However, as it was observed in Ref.\cite{Braaten:1985is}, if  the connection is  compatible
with the metric and the covariant torsion tensor
\bea\label{osaias}
H_{\mu \nu\sigma}=G_{\mu \rho}\ \big(\,{ \Gamma^\rho}_{\nu\sigma}-{ \Gamma^\rho}_{\sigma\nu}\,\big)
\eea
is a closed  three-form: 
\bea\label{sosaoaso}
H_{\mu \nu\sigma}=\partial_\mu B_{\nu\sigma}+\partial_\nu B_{\sigma\mu}+\partial_\sigma B_{\mu\nu}\ ,
\eea
then \eqref{sooisa} follows from
the  Polyakov action with the Kalb-Ramond   term
\bea\label{oisisai}
{\cal A}=
2\, \int_{\Sigma}\rd^2 x\ \Big(\, G_{\mu\nu}\,
\partial_+ X^\mu\partial_- X^\nu-{\textstyle\frac{1}{2}}\
B_{\mu\nu}\
\big(\,\partial_+ X^\mu\partial_- X^\nu-\partial_- X^\mu\partial_+ X^\nu\,\big)\, \Big)\ .
\eea
The 
 2-form $B_{\mu\nu}$  provides an anti-symmetric component 
to the affine connection  and is  sometimes known as  the {\it torsion potential}.
Field theories 
of the  type \eqref{oisisai}  are important in many aspects of physics, from QCD  to condensed matter and
are known as   { Non-Linear Sigma Models} (NLSM). 
The corresponding
Euler-Lagrange equations are usually referred 
to as the   (generalized) {\it harmonic map} problem\ \cite{Fuler:1954}. 
In general, the    NLSM
is a complicated 
structure.
To this order, any simplified example, that softens the sever mathematical problems can
be considered useful and worth studying.

Starting at  the end of the seventies,
an approach was developed
for a  solution of the  generalized harmonic map problem for certain classes of 
the   integrable  target space backgrounds\ \cite{Pohlmeyer:1975nb, Zakharov:1973pp, Eichenherr:1979ci}.
More specifically,  
the term   {\it  integrable}  here  is used to imply that 
Eqs.\eqref{sooisa} constitute
a flatness condition
\bea\label{sososaosa}
[\, {\boldsymbol D}_+(\lambda)\, ,\, {\boldsymbol D}_-(\lambda)\,]=0
\eea
for some matrix valued   worldsheet connection
\bea\label{ssissiaoas}
{\boldsymbol D}_\pm(\lambda)=\partial_\pm+{\boldsymbol A}_{\pm}\ ,\ \ \ \ \ \ \ \ \ \ \ \
{\boldsymbol A}_{\pm}= {\boldsymbol \alpha}^{(\pm)}_\mu(\lambda)\ \, \partial_\pm X^\mu\ ,
\eea
which depends on an arbitrary
complex parameter  $\lambda$.
The approach has been proven to be effective,   especially  concerning  the models  with   homogeneous 
target manifolds $G/H$ and  a  Zero-Curvature Representation (ZCR)
became a central issue 
in  this class of   harmonic map problems \cite{:1989Uhlenbeckci}.
At the same time the integrability of  NLSM  with  non-homogeneous target spaces 
have been  given considerably less attention.
In particular,  there  are   important    examples of  
NLSM which are expected to be  
classical integrable systems despite having   their ZCR     remain unknown.
Among them  are  the   {\it sausage models} 
which were introduced   in Refs.\cite{Fateev:1992tk,Fateev:1996ea}
by taking  advantage of a  perturbative renormalizability   
of a general  NLSM.

The quantum theory governed by the action  \eqref{oisisai}    is  perturbatively renormalizable   and the scale
dependence of its couplings can be computed order by order in perturbation theory\ \cite{Honerkamp:1971sh, Polyakov:1975rr, Friedan:1980jf,
Witten:1983ar, Braaten:1985is, Fradkin:1985ys,  Metsaev:1987bc}.
This, in effect,  
induces deformations of the  metric and the torsion potential with respect to the Renormalization Group (RG)
time $t$, given by (up to the overall factor $\frac{1}{2\pi}$) the logarithm of worldsheet length scale, 
which can be formulated and studied systematically in all
generality. 
The renormalization of the metric and the torsion potential, viewed as generalized couplings, takes the following
form to one-loop\ \cite{ Friedan:1980jf, Braaten:1985is }:
\bea\label{sosasaias}
\frac{\partial}{\partial t}\ \big(\, G_{\mu\nu}+B_{\mu\nu}\,\big)=-\big(\,
{\cal R}_{\mu\nu}+2\, {\cal  D}_\mu V_\nu\,\big)\ ,
\eea
where ${\cal R}_{\mu\nu}$ is the Ricci tensor build from    the  affine connection ${\cal   D}_\mu$.
The
RG equation \eqref{sosasaias} is no other but the {\it Ricci flow} which arose independently in mathematics
(in the  case of   the  Levi-Civita connection) 
as a  tool to address
a variety of non-linear problems in differential geometry and, in particular, the uniformization of compact Riemannian
manifolds\ \cite{:1982Hamilton,:2002Perelman}.
The one-loop RG  equations
\eqref{sosasaias} are highly non-linear and lead to solutions which typically develop singularities.
However, the equations  possess the  {\it ancient} solutions 
which  exist at $t\to-\infty$ and evolve forward in time until the formation of
singularities.\footnote{If  a solution is defined  
for $-\infty<t<+\infty$, it is called an  {\it eternal} solution.} The NLSM underlining the
ancient   solutions have a  good chance to be defined non-perturbatively as a   local integrable quantum field theory.
In the works \cite{Fateev:1992tk}  and  \cite{Fateev:1996ea} 
there were discovered remarkable  ancient   solutions
which describe   torsion-free deformations of the two- and three spheres, respectively (see also Refs.\cite{Bakas:2003xw, Bakas:2009tz}
for comprehensive analysis of these solutions).
The authors conjectured that the solutions describe  the one-loop renormalization of  certain 
quantum field theories  and  performed  highly convincing non-perturbative  analysis  in favor
of their   quantum  integrability.
Note that their  arguments  were based on  the 
$S$-matrix bootstrap  and did not employ any  classical integrable structures.

In this article we attempt to reverse  the logic of Refs.\cite{Fateev:1992tk,Fateev:1996ea} and   apply    the machinery of classical
integrability to produce   ancient   solutions of  the Ricci flow.
We find, as the  main result,  the  ZCR  for  a new four-parameter
family of NLSM with  three-dimensional
target space background.
Note that the proposed ansatz for the ZCR
can be naturally understood in a context   of the averaging procedure 
applied  in the construction of elliptic and trigonometric solutions of the Yang-Baxter
equation from the rational one \cite{Belavin:1981ix, Belavin:1984}.
A similar approach was used in Ref.\cite{Mikhailov:1981} 
to explore  reductions of the Lax
representation (see also  Part II, Chapter IV.2  in  the book \cite{Faddeev:1987ph}).
It turns out that modulo  reparameterizations  encoded by the second term
in the r.h.s. of \eqref{sosasaias}, the effect of  one-loop 
renormalization within  the obtained family
of classical  integrable  NLSM, are
reduced to  the  renormalization of the  parameters
and  the {\it string tension} (the overall  normalization  of the  action).
Therefore, the   family  provides  an interesting example
of multi-parameter
solution of the Ricci flow driven by the connection with torsion.
In the case of a   torsion-free background  the   solution
is reduced to Fateev's  three-dimensional sausage \cite{Fateev:1996ea}.
Since   the  Fateev-Onofri-Zamolodchikov sausage \cite{Fateev:1992tk}
can be obtained  from the three-dimensional one  through a certain limiting procedure,
the    result of this work  yields
the ZCR for   both models.

\section{ZCR for the  harmonic map problem}

Before  focusing   on the case  with  $D=3$ it is useful to      rewrite
Eqs.\eqref{sooisa} in the matrix form for an arbitrary   dimension $D$. 
Suppose  ${\cal M}_D$ is an  oriented manifold and
its   metric can be  transformed   from the coordinate basis to the vielbein one,
$G_{\mu\nu}=e^{a}_\mu\,  e^{a}_\nu$.
Introduce  the conventional  one-forms acting in  a spinor representation  of $SO(N)$, namely
 the Levi-Civita spin connection  ${\boldsymbol \omega}_\mu$ and
 ${\boldsymbol \gamma}_\mu= \gamma^a\, e^a_\mu$,
where  $\gamma$-matrixes obey  the standard Dirac algebra,
$\{\gamma^a,\,\gamma^b\,\}=2\, \delta^{ab}$. 
With the use of these notations the generalized  harmonic map problem \eqref{sooisa} can  be  brought to
the form
\bea\label{sposaoasoas}
\big[\, \partial_+ +  {\boldsymbol\omega}^+_\mu\,\partial_+X^\mu,\,
{\boldsymbol \gamma}_\nu\,\partial_-X^\nu\,\big]=0\, ,\ \ \ \ \ \ \
\big[\,{\boldsymbol \gamma}_\nu\,\partial_+X^\nu\,,\,
 \partial_-+  {\boldsymbol\omega}^-_\mu\,\partial_-X^\mu\,\big]=0\ ,
\eea
where
${\boldsymbol {\omega}}^\pm _\mu={\boldsymbol {\omega}}_\mu\mp 
{ \frac{1}{8}}\ H_{\mu\nu\sigma}\ 
{\boldsymbol \gamma}^\nu{\boldsymbol \gamma}^\sigma$
and as usual, ${\boldsymbol {\gamma}}^\mu=G^{\mu\nu}\,  {\boldsymbol {\gamma}}_\nu$.

For $D=3$,
$\gamma$-matrixes can be identified with the conventional Pauli matrixes
$\gamma^a=\sigma_a, \  a=1,\,2,\,3$, whereas 
$H_{\mu\nu\sigma}$
must be proportional to the
volume form,
\bea\label{aosaisia}
H_{\mu\nu\sigma}={\rm H}\   \sqrt{G}\ \ \  { \frac{1}{3!}}\ \epsilon_{\mu\nu\sigma}\ .
\eea
In this case, the matrix valued one-forms ${\boldsymbol {\omega}}^\pm$ in \eqref{sposaoasoas} is simplified  to
\bea\label{osaoasaso}
{\boldsymbol \omega}^\pm_\mu={\boldsymbol \omega}_\mu\mp \frac{\ri}{4}\  {\rm H}\ {\boldsymbol \gamma}_\mu\ .
\eea
Without making any symmetry assumptions, 
the classification of NLSM  possessing the  zero-curvature representation 
seems to be a hopeless task even for  lower dimensional target manifolds.
We are therefore forced to impose some symmetry conditions on the metric and
the torsion potential. Let us  assume that the target  space background possesses two
commuting Killing vector fields. More specifically, there exist 
a local coordinate frame    $X^\mu=(u, v, w)$ with respect to which
the  metric 
takes the form
\bea\label{opososa}
G_{\mu\nu}\,\rd X^\mu\rd X^\nu =G_{uu}\ (\rd u)^2+G_{vv}\ (\rd v)^2+\ G_{ww}\ (\rd w)^2+2 \, G_{vw}\ \rd v  \rd w\ ,
\eea
and   the components of the metric tensor, as well as the torsion strength  ${\rm H}$ in  \eqref{aosaisia}, do not depend on
the coordinates $v$ and $w$.  
Without further  loss of generality we   set
$\sqrt{G_{uu}}$  to be a positive constant
\bea\label{sososasq}
\sqrt{G_{uu}}=const>0\ .
\eea
Given  the metric $G_{\mu\nu}$, the introduction of the tangent vectors  $e^a_\mu$
involves arbitrary choices at each point of ${\cal M}_3$.
We are free to make local $SO(3)$ rotation  on the index $a$, or equivalently adjoint
 $SU(2)$ transformation  on ${\boldsymbol {\gamma}}^\mu$.
The transformation law of  the 1-form  ${\boldsymbol \omega}^\pm_\mu$ includes an inhomogeneous piece typical 
of gauge fields,
\bea\label{sososaoare}
{\boldsymbol \gamma}_\mu\to {\boldsymbol U}^{-1}\,
 {\boldsymbol \gamma}_\mu\,
{\boldsymbol U}\ ,\ \ \ \ \
{\boldsymbol \omega}^\pm_\mu\to {\boldsymbol U}^{-1}\,
 {\boldsymbol \omega}^\pm_\mu\, {\boldsymbol U}+ {\boldsymbol U}^{-1}\partial_\mu
{\boldsymbol U}\ .
\eea
For the metric of the form\ \eqref{opososa} the  gauge freedom can be used to
set  $e^1_u=e^2_u=e^3_v=e^3_w=0$.  Then the
non-vanishing components of the vielbein are  defined by the  formulas
\bea\label{sosaoasi}
G_{uu}=(e_u^3)^2\, ,\ \ \ \ \ 
G_{vv}=e_v^{+} e_v^{-}\, ,\ \  \ \ \ \
G_{ww}=e_w^{+} e_w^{-}\, ,\ \  \ \ \ \
G_{vw}={\textstyle \frac{1}{2}}\ ( e_v^{+} e_w^{-}+e_v^{-} e_w^{+})
\eea
modulo       $U(1)$  rotations $e^{\pm}_\mu\to \re^{\pm\ri\phi} \ e^{\pm }_\mu$,  
where  
\bea\label{aasasio}
e^{\pm}_\mu=e^1_\mu\pm \ri\, e^2_\mu
\eea
and $\phi=\phi(u)$ is an arbitrary local phase.
Below we use the fact that
the residual  freedom in the    choice  the vielbein  implies
the local    symmetry    \eqref{sososaoare},  where ${\boldsymbol U}$ is substituted by the diagonal
matrix
\bea\label{isusosasasai}
{\boldsymbol U}_\phi=
\exp\big(\,{\textstyle\frac{\ri}{2}}\, \phi(u)\,\sigma_3\,\big)\ .
\eea
Note that the sign of $e^3_u=\pm\sqrt{G_{uu}}$ is actually unambiguous
for the  chosen   orientation of  the vielbein (i.e.,  for  the chosen  sign of $\sqrt{G}:=\det(e^a_\mu)$).

We turn now to the construction of  the ZCR.
Let ${\boldsymbol \zeta}_\mu(\lambda)$  be a  matrix valued 1-form  which depends on the spectral parameter
\bea\label{sopsosaosa}
{\boldsymbol \zeta}_\mu(\lambda)=\sum_{a=\pm, 3} { f}_a(\lambda)\, e^a_\mu\ \sigma_a \ .
\eea
Here $\sigma_\pm=\frac{1}{2}\, (\sigma_1\mp\ri \sigma_2)$ and 
${ f}_a(\lambda)$
read explicitly as follows
\bea\label{sosapsaosap}
f_+( \lambda)&=&-f_-(-\lambda)=
\frac{1}{\sqrt{G_{uu}} }\ \  \frac{\vartheta_1(u-\frac{\lambda}{2}, q)\,
\vartheta'_1(0, q)}
{2\ri\, \vartheta_1(\frac{\lambda}{2}, q)\,
\vartheta_1(u, q)}\nonumber\\
f_3(\lambda)&=&\frac{1}{\sqrt{G_{uu}}}\ \
\frac{\vartheta_1'(\frac{\lambda}{2}, q)}
{2\ri\, \vartheta_1(\frac{\lambda}{2}, q)}\ .
\eea
In the l.h.s. of the  above  equations, we  only indicate  dependence on
the spectral parameter, $\vartheta_1$ stands for   the conventional theta function of the nome 
$q=\re^{\ri\pi\tau}$  $(0<q<1)$  and 
$\vartheta'_1(u, q):=\partial_u \vartheta_1(u, q)$.
To get a more informal feel for  ${\boldsymbol \zeta}_\mu$,
let us note that
it can be alternatively  defined through the principal value summation of the formal 
double series
\bea\label{sososaioas}
{\boldsymbol \zeta}_\mu(\lambda)= \frac{1}{ \sqrt{G_{uu}}}\ \  V.P. \sum_{n,m=-\infty}^\infty
 \frac{1}{\ri}\ \frac{\re^{\ri nu\sigma_3}\ {\boldsymbol \gamma}_\mu\ \re^{-\ri nu\sigma_3}}
{\lambda+2\pi (m+n\tau)}\ .
\eea 
Using  ${\boldsymbol \zeta}_\mu$ as a building block, we define
the worldsheet connection of  the form\ \eqref{ssissiaoas} with
\bea\label{soosaisa}
{\boldsymbol \alpha}^{(+)}_\mu(\lambda)={\boldsymbol \alpha}_\mu(\lambda\,|\,\eta_+,\phi_+)\ ,
\ \ \ \ \ \ 
{\boldsymbol \alpha}^{(-)}_\mu(\lambda)={\boldsymbol \alpha}_\mu(\lambda-\pi\,|\,\eta_-,\phi_-)\ ,
\eea
and
\bea\label{tsososa}
{\boldsymbol \alpha}_\mu(\lambda\, |\, \eta, \phi)=
\frac{1}{2\ri }\ \Big(\, {\boldsymbol U}_\phi^{-1}\  {\boldsymbol \zeta}_\mu(\ri\,\eta+\lambda)\, 
{\boldsymbol U}_\phi+
\sigma_2\ {\boldsymbol U}_\phi^{-1}\   {\boldsymbol  \zeta}_\mu(\ri\,\eta-\lambda)\,
{\boldsymbol U}_\phi\ \sigma_2\,\Big)\ .
\eea
Here  $\eta_+$ and $\eta_-$ stand for  arbitrary  parameters whereas $\phi_\pm=\phi_\pm(u)$ are 
arbitrary local phases showing up in the matrices of the form  \eqref{isusosasasai}.
The local twists are included  in  \eqref{tsososa}
because of the  residual  freedom in the    choice  the vielbein.

Under these definitions, it is  straightforward  to establish the following properties:
\begin{itemize}

\item {\it Quasiperiodicity}.
\bea\label{saososaisa}
{\boldsymbol D}_\pm(\lambda)=
\re^{\ri n u\sigma_3}\  {\boldsymbol D}_\pm\big(\lambda+2\pi
\, (m+n\tau)\,\big)
\ \re^{-\ri n u\sigma_3}\ \ \ \ \ \ \ (\,m,\,n\in {\mathbb Z}\,)\ .
\eea
\item {\it $\lambda$-parity}.
\bea\label{ososao}
{\boldsymbol D}_\pm(-\lambda)=\sigma_2\ {\boldsymbol D}_\pm(\lambda)\, \sigma_2\ .
\eea
\item {\it Singularities}.
Let $|\Im  m(\eta_+)|< \pi\,,\ |\Re e(\eta_+)|<   \Im m(\pi\tau)$. 
In 
the fundamental parallelogram $(-\pi,\,\pi )\otimes (-\pi \tau,\,\pi\tau)$,  
${\boldsymbol D}_+(\lambda)$
has two simple poles
with the residues
\bea\label{soisaisa}
{\boldsymbol D}_+(\lambda)&=&\frac{1}{\sqrt{G_{uu}}}\ \ \frac{1}{2(\lambda+\ri\eta_+ )}\ \  {\boldsymbol U}^{-1}_{+}\,
{\boldsymbol \gamma}_\mu \, 
{\boldsymbol U}_{+} \ \partial_+X^\mu+O(1)\\
&=&\frac{1}{\sqrt{G_{uu}}}\ \ \frac{1}{2(\lambda-\ri\eta_+)}\ \ \sigma_2\, 
{\boldsymbol U}^{-1}_{+}\,{\boldsymbol \gamma}_\mu\,
{\boldsymbol U}_{+}\, \sigma_2\ \partial_+X^\mu+O(1)\ . \nonumber
\eea
Similarly, the singularities of   ${\boldsymbol D}_-(\lambda)$   in
the parallelogram  $(0 ,\, 2\pi)\otimes (-\tau,\tau)$
are given by
\bea\label{soisaisar}
{\boldsymbol D}_-(\lambda)&=&-\frac{1}{\sqrt{G_{uu}}}\ \ \frac{1}{2(\lambda-\pi+\ri\eta_-)}\ \  {\boldsymbol U}^{-1}_{-}\,
{\boldsymbol \gamma}_\mu \, 
{\boldsymbol U}_{-} \ \partial_-X^\mu+O(1)\\
&=&\frac{1}{\sqrt{G_{uu}}}\ \ \frac{1}{2(\lambda-\pi-\ri\eta_-)}\ \ \sigma_2\, 
{\boldsymbol U}^{-1}_{-}\,{\boldsymbol \gamma}_\mu\,
{\boldsymbol U}_{-}\, \sigma_2\ \partial_-X^\mu+O(1)\ . \nonumber
\eea
Here we use the shortcut notations ${\boldsymbol U}_{\pm}={\boldsymbol U}_{\phi_{\pm}}$.

\item {\it Hermiticity.} 
Let $0<q<1$, $\Im m (\eta_\pm)=0,\  |\eta_\pm|<   \Im m(\pi\tau)$, then
\bea\label{saossaop}
{\boldsymbol D}^\dagger_\pm(\lambda)=-{\boldsymbol D}_\pm(-\lambda^*)\ .
\eea
In particular,
${\boldsymbol D}_\pm$  are (formally) anti-Hermitian differential
operators as $\Re e(\lambda)=\pi n$ $(n=0,\,\pm1\ldots)$.

\end{itemize}
The first two   properties of the worldsheet  connection  imply that the field strength
${\boldsymbol F}(\lambda)=[{\boldsymbol D}_+(\lambda), {\boldsymbol D}_-(\lambda)]$ satisfies the conditions:
\bea\label{aososaisa}
{\boldsymbol F}(\lambda)&=&
\re^{\ri n u\sigma_3}\  {\boldsymbol F}\big(\lambda+2\pi
\, (m+n\tau)\,\big)
\ \re^{-\ri n u\sigma_3}\ \ \ \ \ \ \ (\, m,\,n\in {\mathbb Z}\,)\nonumber\\
{\boldsymbol F}(\lambda)&=&\sigma_2\,  {\boldsymbol F}(-\lambda)\,\sigma_2\   .
\eea
We may  try to adjust the target space background to make $2\times 2$ matrix   
${\boldsymbol F}(\lambda)$ nonsingular 
in the whole complex plane of $\lambda$.
Using the matrix form \eqref{sposaoasoas} of the harmonic map equations, it is easy to see that 
the cancellation of the poles  of ${\boldsymbol F}(\lambda)$ yields the
relations
\bea\label{spososaojsh}
{\boldsymbol \omega}^+_\mu&=&
{\boldsymbol U}_+\, {\boldsymbol \alpha}_\mu(\pi-\ri\eta_+\, |\, \eta_-, \phi_-)\,
{\boldsymbol U}^{-1}_{+}+{\boldsymbol U}_+\, \partial_\mu\, {\boldsymbol U}^{-1}_{+}\\
{\boldsymbol \omega}^-_\mu&=&
{\boldsymbol U}_-\, {\boldsymbol \alpha}_\mu(\pi -\ri\eta_-\, |\, \eta_+, \phi_+)\,
{\boldsymbol U}^{-1}_{-}+{\boldsymbol U}_-\, \partial_\mu\, {\boldsymbol U}^{-1}_{-}\ ,\nonumber
\eea
or, equivalently,
\bea\label{ossasaus}
{\boldsymbol \omega}^\pm_\mu&=&\frac{1}{2\ri}\, \Big[\,
\sigma_2\ ({\boldsymbol U}_+{\boldsymbol U}_-)^{-1}
 {\boldsymbol \zeta}_\mu\big(\,\pi+2 \ri\eta)\,\big)\, {\boldsymbol U}_+{\boldsymbol U}_-\ 
\sigma_2\nonumber\\
&+& {\boldsymbol U}_\pm {\boldsymbol U}_\mp^{-1}
\, {\boldsymbol \zeta}_\mu\big(\,\pi\mp 2\ri\nu
\,\big)\, {\boldsymbol U}_\mp{\boldsymbol U}_\pm^{-1}
+2\ri \ {\boldsymbol U}_\pm\, \partial_\mu\, {\boldsymbol U}^{-1}_\pm\, \Big]\  ,
 \eea
where $\eta$ and $\nu$ stand for
\bea\label{sosoosa}
\eta=\frac{\eta_++\eta_-}{2}\ ,\ \ \ \ \nu=\frac{\eta_+-\eta_-}{2}\ .
\eea
If we proceed further and impose an  extra  condition
\bea\label{siosisasia}
{\boldsymbol F}(0)=0\ ,
\eea
then 
${\rm Tr}[\,{\boldsymbol F}^2(\lambda)\,]$ becomes an  entire,
doubly  periodic function of $\lambda$, vanishing at $\lambda=0$. Hence
it must be identically zero. Combining this fact  with  the  hermiticity we find  that
the field strength vanishes
for any pure  imaginary $\lambda$: ${\boldsymbol F}(\lambda)=0\, , \Re e(\lambda)=0$.
Of course, this implies that  the conditions
\eqref{spososaojsh} and \eqref{siosisasia}  guarantee
the flatness of the worldsheet  connection.

Let us take a closer look at the condition \eqref{siosisasia}. Because of the 
$\lambda$-parity relation \eqref{ososao},
the connection reduces at $\lambda=0$  to the form
\bea\label{iissaio}
{\boldsymbol D}_\pm(0)=\partial_\pm\mp \ri\ I_{\pm}\ \sigma_2\ .
\eea
Therefore, the flatness condition  implies a continuity equation $\partial_+I_-+
\partial_-I_+=0$.
For the target space background with  the two
Killing vector fields $\frac{\partial}{\partial v}$ and $\frac{\partial}{\partial w}$
the NLSM possesses two Noether currents,  $V_{A}$ and $W_{A}$.
(Here we label the   worldsheet    components  by the  subscript $A=\pm$\,.)
Thus, we may  conclude that the flatness condition at $\lambda=0$  \eqref{siosisasia} is equivalent to the
relation
\bea\label{sososaysts}
I_{A}=c_v\ V_{A}+c_w\ W_{A}\ ,
\eea
where $c_v$ and $c_w$  are  some   real constants.

The conditions \eqref{ossasaus} and \eqref{siosisasia} can be treated as a set of equations
for the determination of the non-vanishing vielbein components,  the torsion strength ${\rm H}$ and 
the unknown  phases $\phi_\pm$.
It is easy to see without actually doing any computation that 
the solution, if it exists,  is not unique.
Indeed,
under the diffeomorphism  $X^\mu\to {\tilde X}^\mu$ the vielbein transforms as
${e^a_\mu}= {{\tilde e}^a_\nu}\ \frac{\partial {\tilde X}^{\nu}}{\partial X^\mu}$, therefore the
$SL(2,R)$ coordinate
transformations with a  constant Jacobian matrix
\bea\label{osososa}
\frac{\partial {\tilde X}^{\mu}}{\partial X^\nu}=
\begin{pmatrix}
1&0&0\\
0&S_{2}^2& S_{2}^3\\
0&S_{3}^{2}&S_{3}^3\\
\end{pmatrix}\ ,\ \ \ \ \ \ \ \ \ \ \ \ \ \ S_{2}^{2}S_{3}^{3}-S_{2}^{3}S_{3}^2=1\ ,
\eea
preserve the torsion strength and    the  general form  of the metric\ \eqref{opososa}.
These   transformations  can be applied to generate
a three parameter family
from any  given  solution of  Eqs.\eqref{ossasaus} and \eqref{siosisasia}.
It is also clear  that
the equations   impose   restrictions on the
phase difference  $\phi_+-\phi_-$ only, i.e., one out of two  phases can be chosen at will.
In  Appendix we found the most general solution of  Eqs.\eqref{ossasaus}.
It turns out that using the $SL(2,R)$ transformations \eqref{osososa} 
and the orientation-preserving transformation  $(u, v,w)\to (-u,w,v)$, 
the solution can be brought to the form
\bea\label{uissasaus}
e^3_u&=& g^{-1}\
l\ {\varepsilon}(u)\nonumber\\
e^\pm_v&=&g^{-1}\ 
\re^{\pm \ri(\phi_+-\frac{\pi}{2})}\  \rho(\pm u)\
\frac{{\vartheta}_{4}(\ri\eta\pm u, q^2)}{{\vartheta}_{4}(\ri\eta, q^2)}
\\
e^\pm_w&=&-g^{-1}\
\re^{\pm \ri(\phi_+-\frac{\pi}{2})}\  \rho(\pm u)\
  \frac{{\vartheta}_{1}(\ri\eta\pm u,q^2)}
{{\vartheta}_{1}(\ri\eta, q^2)}\nonumber
\eea
and
\bea\label{ssiosisosa}
&&{\rm H}=\frac{g}{l}\ \bigg[\, \ri\  \frac{\vartheta'_2(\ri\nu,q)}{\vartheta_2(\ri\nu,q)}-\ri\
\partial_u\log\bigg(\frac{{ \rho}(u)}{{\rho}(-u)}\bigg)\, \bigg]\nonumber\\
&&\re^{\ri (\phi_+-\phi_-)}=\frac{\rho(u)}{\rho(-u)}\ .
\eea
Here we use the notation
\bea\label{kksisisa}
\rho(u)=\frac{
\vartheta_3(\frac{\ri\sigma-\ri\nu}{2},q)\vartheta_4(\frac{\ri \sigma+\ri\nu}{2},q)}
{\vartheta_3(\frac{u+\ri\sigma-\ri\nu}{2},q)\vartheta_4(\frac{u-\ri \sigma-\ri\nu}{2},q)}\ ,
\eea
and $ \varepsilon(u)$ is a periodic step function
\bea\label{aisasoix}
 \varepsilon(u)=
\begin{cases}
+1\ ,\ \ \ \ \ \ u/\pi\in (\, 2 n ,\,  2n+1\,) \\
-1\ ,\ \ \ \ \ \ u/\pi\in ( 2 n+1,\, 2n+2\, )
\end{cases}\
\ \ \ \ \ (n\in {\mathbb Z})\ .
\eea

The above formulas  call for a number of comments.
The constant 
$g>0$ (the string tension)  just sets  the overall normalization for the NLSM action\ \eqref{oisisai},
which does not affect the Euler-Lagrange   equations.
It can be set to be  one without loss of any generality.
We however  reserve this parameter  and make use of it later for the purpose of   quantization.
The  parameter   $l$ in fact just replaces   $\sqrt{G_{uu}}$.
It can  be absorbed into $g$ and  the overall normalization of
the Killing coordinates $v$ and $w$.
It is convenient to choose it as
\bea\label{issusa}
l=\ri\ \frac{\vartheta_2(\ri\eta,q)\,\vartheta_1'(0,q)}
{\vartheta_1(\ri\eta,q)\,\vartheta_2(0,q) }\ .
\eea
Then, as it follows from\ \eqref{uissasaus}
\bea\label{sosoasa}
\sqrt{G}=
\frac{l^2}{g^3}\ \ \rho( u) \rho(- u)\ \
\ \frac{ \vartheta_1(u,q)}{\vartheta'_1(0,q)}\ \varepsilon(u)\ .
\eea
Since under the complex conjugation,   $\rho^*(u)=\rho(-u)$,
the periodic step function $\varepsilon(u)$ provides the positivity
of  $\sqrt{G}=\det(e^{a}_\mu)$
for 
an arbitrary real  $u$ except
$u=0,\pm \pi,\,\pm2\pi\ldots\ .$
Thus     nontrivial parameters of the solution are $q, \eta, \nu$, which have already appeared in 
the ansatz for the worldsheet connection,
and    $\sigma$ from the  definition\ \eqref{kksisisa}.

It turns out
that the condition\ \eqref{siosisasia}
does not
impose any   restriction on
the above solution. More specifically,   through proper choice of the constants
$c_v$ and $c_w$,
Eq.\eqref{sososaysts} is identically   satisfied.

\section{The integrable target space background}

In this section we would like to  discuss  the integrable
target space background, i.e., the metric
and the torsion potential
in the NLSM action\ \eqref{oisisai}.
We start with  the  $q\to 0$ limit, assuming
all other parameters are held constant. It is easy to see that
\bea\label{taraers}
&&G_{\mu\nu}\rd X^\mu\rd X^\nu=
{\frac{1}{g^2}}\, \Big[\, C^2\, (\rd u)^2+   
(\rd v)^2+\big(1+(C^2-1) \sin^2(u)\big)\, (\rd w)^2 -2\cos(u)\, \rd v\rd w
\,\Big]\nonumber\\
&&
\frac{1}{2!}\
B_{\mu\nu}\,\rd X^\mu\wedge \rd X^\nu=
\frac{  C\tanh(\nu) }{g^2}\ \ \cos(u)\  \rd v\wedge\rd w\ \ \ \ \ \ (q=0)\ ,
\eea
where $C=\coth(\eta)$.  Note that the parameter $\sigma$ does not appear in this limiting  case.
We may now send $\eta\to+\infty$:
\bea\label{ussaias}
&&G_{\mu\nu}\,\rd X^\mu\rd X^\nu=
\frac{1}{g^2}\ \Big[\,  (\rd u)^2+ 
(\rd v)^2+(\rd w)^2 -2\cos(u)\, \rd v\,\rd w\,
\Big]\\
&&\frac{1}{2!}\
B_{\mu\nu}\,\rd X^\mu\wedge \rd X^\nu=
\frac{\tanh(\nu)}{g^2}\ 
\cos(u)\ \rd v\wedge\rd w
\ \ \ \ \ \ \ \ (q=0,\ \eta\to+\infty)\ .\nonumber
\eea
Let ${ \vec n}$ be a unit vector in  ${\mathbb R}^4$ whose  components are defined
by the relations
\bea\label{saoisisa}
n^1\pm \ri \,n^2=\re^{\pm \frac{\ri (v-w)}{2}}\ \sin\big({\textstyle\frac{u}{2}}\big)\ ,
\ \ \ \ \ \ \ n^3\pm \ri \,n^4=\re^{\pm \frac{\ri (v+w)}{2}}\ \cos\big({\textstyle\frac{u}{2}}\big)\ ,
\eea
then 
the metric \eqref{ussaias} takes  the form
$\frac{4}{g^2}\,
 \rd {\vec n}\cdot\rd {\vec n}$,
i.e. it  coincides   with  the round
metric on the three-sphere of radius $\frac{2}{g}$.\footnote{Note the slightly non-standard 
choice for the coupling  $g$.}
The variables 
$\theta=\frac{u}{2}$, $\chi_1=\frac{v- w}{2}$ and  $\chi_2=\frac{v+ w}{2}$ are usually referred as to
the Hopf coordinates and used in the description of the three-sphere as
the Hopf bundle.
For any value of $\theta$ between $0$ and $\frac{\pi}{2}$,  the pair
$(\chi_1, \,\chi_2)$ parameterizes a two-dimensional torus $(\chi_a\sim\chi_a+2\pi,\ a=1,2)$.
The metric \eqref{ussaias}   degenerates  (i.e.  $\sqrt{G}=0$) at  two sub-manifolds of codimension two (circles)
which  correspond to $u=0$ and $u=\pi$.
However 
these are coordinate singularities that may be removed by  introducing   suitable coordinates.
The same happens for the  one-parameter family of metrics \eqref{taraers} which is 
sometimes referred as to
a  metric on  a squashed 
three-sphere.\footnote{The NLSM with the target space background\ \eqref{taraers} is  called  the  anisotropic   $SU(2)$ Wess-Zumino-Witten-Novikov
model.
The ZCR  for this  model  is known for a while  (see\cite{Zakharov:1973pp},\, \cite{Cherednik:1981df},\,
Chapter II.1.5 in   the  monograph  \cite{Faddeev:1987ph}
and references therein).}

In order to give an  explicit  description of the general target space background
it is convenient to replace the
coordinate $u$ with another variable.
As  it follows from
Eq.\eqref{sosoasa} the metric  degenerates at $u=0,\,\pm\pi,\,\pm2\pi\ldots\ .$ Suppose
the coordinate $u$ runs over the segment
\bea\label{soissai}
0<u<\pi\ ,
\eea
then the doubly periodic function ${z}={z}(u,q)$,
\bea\label{sossaius}
{z}(u,q)=
\frac{\vartheta_{2}(u, q^2) \vartheta_{3}(0, q^2)}{\vartheta_{3}(u, q^2)
\vartheta_{2}(0, q^2)}\ ,
\eea
varies monotonically along the  segment $(-1,1)$
and therefore $u$  can  be replaced by  $z$.
Introduce 
\bea\label{ksisusossai}
p=-\ri\ \frac{\vartheta_{1}(\ri\eta,q^2)}{\vartheta_{4}(\ri\eta,q^2)}\ ,\ \ \ \ \ \ \ \
h=-\ \ri\ \sqrt{\kappa}\ \ \frac{\vartheta_{1}(\ri\nu-\ri\sigma ,q^2)}{\vartheta_{4}(\ri\nu-\ri\sigma,q^2)}\ ,\ \ \ \
{\bar h}=-\ri\ \sqrt{\kappa}\ \ \frac{\vartheta_{1}(\ri\nu+\ri\sigma,q^2)}{\vartheta_{4}(\ri\nu+\ri\sigma,q^2)}
\eea
and
\bea\label{ystsosoasi}
\sqrt{\kappa}=\frac{\vartheta_2(0,q^2)}{ \vartheta_3(0,\,q^2)}\ ,
\eea
which can be  viewed   as  a  new set of  parameters replacing $(\eta,\nu,\sigma, q)$. 
To make  formulas  more readable, we  will  also use  the following     combinations of the  new parameters:
\bea\label{psooisisaais}
&&c=+\sqrt{\frac{1+ h^2}{\kappa^2 + h^2}}\ ,\  \ \ \ \ \ \
{\bar c}=+\sqrt{\frac{1+{\bar h}^2}{\kappa^2 +{\bar h}^2}}\\
&&m=+\sqrt{1+\kappa^2+\kappa p^2+\kappa p^{-2}}\ .\nonumber
\eea
With the new coordinate frame $(z,v,w)$ and the new set of parameters $(p, h,{\bar h}, \kappa)$,
the metric  defined by Eqs.\eqref{opososa},\,\eqref{sosaoasi},\,\eqref{uissasaus},\,\eqref{issusa}
can be  brought to the form:
\bea\label{isussossais}
&&G_{\mu\nu}\ \rd X^\mu\rd X^\nu=\frac{m^2}{g^2}\ 
\frac{   (\rd z)^2   }{(1-z^2)(1-\kappa^2\, z^2)}+
\frac{1}{g^2}\     
\frac{(c+1)({\bar c}-1)}{(1-\kappa^2)(c+z)( {\bar c}-z)}\times\nonumber\\
&&\ \ \ \Big[\,  \big(\, 1+\kappa\, p^2- z^2\ \kappa\, (\kappa+ p^2)
\,\big)\  (\rd v)^2+
\big(\, 1 +\kappa\, p^{-2}- z^2\ \kappa\, (\kappa+\, p^{-2})
\,\big)\  (\rd w)^2
\nonumber\\
&&\ \ \ -2\ (1-\kappa^2)\ z\ \rd v\,\rd w\ \Big]\ ,
\eea
whereas the torsion potential ${\rm B}$
$\big(\,\frac{1}{2}\, B_{\mu\nu}\,\rd X^\mu\wedge \rd X^\nu=  {\rm B}\, \rd v\wedge \rd w\,\big)$
and torsion strength  ${\rm H}$    $(\partial_u {\rm B}=\sqrt{G}\ {\rm H})$
are given by
\bea\label{ystssosais}
{\rm B}&=&-\frac{m    }{g^2}\ \frac{ (c+1)({\bar c}-1)}{ (1-\kappa^2) ({\bar c}+c)}\    (1- z)\
\  \bigg[\, h\  \frac{c-1}{c+z}+{\bar h}\  \frac{{\bar c}+1}{{\bar c}- z}\, \bigg]\\
{\rm H}&=&\frac{g}{ m}\   \ \
\frac{ h\ ( c^2-1) ({\bar c}- z)^2+{\bar h}\ ({\bar c}^2-1)(c+ z)^2}
{({\bar c}+c)( c+ z) ({\bar c}- z)}\ .\nonumber
\eea

A few comments are in order here. Although
the  parameters $h$ and ${\bar h}$  appear in the metric   through the combinations $c$ and ${\bar c}$ only, 
there are two reasons  to  choose     $(h, {\bar h})$ as  independent parameters.
First,  $h$ and ${\bar h}$ are  fully unrestricted real
numbers, i.e.,
\bea\label{sosisuias}
-\infty<h,\,{\bar h}<+\infty\ ,
\eea
whereas $1<c\leq \kappa^{-1}$ and  $1< c\leq \kappa^{-1}$. Second,
Eqs.\eqref{psooisisaais} allow one  to express $(h, {\bar h})$ through  $(c,{\bar c})$
modulo  sign factors requiring special care
since the torsion potential  substantially  depends on the  relative  sign of $h$  and $ {\bar h}$.
Note the neither metric nor the torsion potential depends on  a sign of $p$. Moreover
the form of   the target space background are  invariant with respect to the  interchange
$v\leftrightarrow w$ accompanied by the transformation of the parameters $(p,h,{\bar h})\to(p^{-1},-h,-{\bar h})$.
Therefore, the actual parameter is $P^2$,
\bea\label{paosaos}
P=\frac{1}{2}\ (\,p-p^{-1}\,)\, , 
\eea 
rather than $p$ itself. We  keep the notation $p$  to make  the equations easier to visualize.
Finally, the parameter  $\kappa$ in \eqref{ystsosoasi} is no other but 
the  elliptic modulus associated with the elliptic nome $q^2$, therefore
\bea\label{saosaosai}
0\leq \kappa< 1\ .
\eea

It is instructive to  look at    the metric \eqref{isussossais} at the degeneration  points $z=z_a\ (a=1,\,2)$  
which correspond to  $u=0,\,\pi$.
We introduce two  
different   local coordinate frames  $(\rho_1 ,\psi_1,\chi_1)$
and  $(\rho_2 ,\psi_2, \chi_2)$ in the vicinity of $z_1=+ 1$ and $z_2=-1$, respectively.
Let
$\rho^2_a=2\, |z-z_a|+O(|z-z_a|^2)$,
\bea\label{sososa}
\psi_1={\frac{1}{2}}\ \big(\, (1+\Delta)\,  v+ (1-\Delta)\, \,w\,\big)\ ,
\ \ \ \ \ \ \ \ \ \  \ \psi_2=
{\frac{1}{2R}}\ \big(\, (1+\Delta)\,  v- (1-\Delta)\, \,w\,\big)
\eea
and
\bea\label{sosssiuaus}
\chi_1=\frac{1}{2 R}\ (v-w)\ ,\ \ \ \ \ \ \ \chi_2=\frac{1 }{2}\ (v+w)\ ,
\eea
where $\Delta=\frac{\kappa}{m^2}\  (p^2-p^{-2})$,
\bea\label{sosisa}
R=\sqrt{\frac{(c-1)({\bar c}+1)}{(c+1)({\bar c}-1)}}\ .
\eea
Evaluating the metric
relative to the  local coordinate systems 
$(\rho_1 ,\psi_1,\chi_1)$ and $(\rho_2 ,\psi_2, \chi_2)$, one finds
\bea\label{sososaos}
G_{\mu\nu}\ \rd X^\mu\rd X^\nu=C^{(1)}_a\ 
\Big(\,(\rd \rho_a )^2+\rho^2_a\, (\rd \psi_a)^2+O\big(\rho^4_a \big)\,\Big)+  
C^{(2)}_a\ (\rd \chi_a)^2\, \big( 1+
O(\rho^2_a)\,\big)\, ,
\eea
where $\rho_a\to 0$   $(a=1,2)$ and
$C_{a}^{(1,2)}$ stand  for  some positive constants. This general  form implies that to avoid
the formation of  the conical singularities at $\rho_1=0$ and $\rho_2=0$, 
the local coordinates $\psi_1$ and $\psi_2$ have to  be  the   angular type variables such that
$\psi_a\sim\psi_a+2\pi$.

In fact, we did not make any assumptions
about global properties of the Killing  coordinates $v$ and $ w$ 
in the derivation of the ZCR.  Therefore, it remains valid for 
any  compactification of these variables.
In what follows  we will assume that
\bea\label{usooiasi}
\chi_1\sim\chi_1+2\pi\ ,\ \ \ \chi_2\sim\chi_2+2\pi\ ,
\eea
where $\chi_a$ are given  by Eq.\eqref{sosssiuaus}. 
In this case the ``global'' chart   
\bea\label{isosasai}
\big(\,z,\,\chi_1,\,\chi_2\,|\, -1<z<1,\  0\leq\chi_{a}<2\pi\,\big)
\eea 
covers  the whole  target space ${\cal M}_3$  except  two  sub-manifolds   of codimension two.
The sub-manifolds are  circles
parameterized  by the angular   variables  
$\chi_1$ as $z=1$ and $\chi_2$ as  $z=-1$. 
Let us consider the neighborhoods of the circle at $z=1$.
We need at least two local charts  with $\chi_1\in (a,b)$ and  $0<b-a<2\pi$ to cover the circle completely.
As it follows from Eqs.\eqref{sososa}, \eqref{sosssiuaus}
\bea\label{iosisa}
\psi_1=\chi_2+R\,\Delta\ \chi_1\ ,\ \ \ \ \ \ \ \ \psi_2=\chi_1+ R^{-1}\Delta\ \chi_2\ ,
\eea
and hence,  at a  local chart with  the decompactified coordinate $\chi_1$, 
the variable
$\psi_1$ is  of the   angular type  provided  the compactification condition \eqref{usooiasi} is imposed.
A similar analysis can be applied to the  neighborhoods of the circle at $z=-1$.

To summarize,  the formula \eqref{isussossais} supplemented by the  global conditions
\eqref{sosssiuaus},\,\eqref{usooiasi},
defines a nonsingular  metric on a  topological   three-sphere  ${\cal M}_3$.

An important integral characteristic of the target space background is a ${\cal H}$-flux, i.e., 
a total   flux  of the
closed three-form ${\cal H}=H_{\mu\nu\sigma}\,\rd X^\mu\wedge\rd X^\nu\wedge\rd X^\sigma$ through the
target manifold. Bellow we will use
\bea\label{asaisosaios}
N=\frac{1}{16\pi^2}\ \int_{{\cal M}_3}{\cal  H}\ .
\eea
With  Eq.\eqref{ystssosais} and the  compactification conditions\ \eqref{usooiasi} one finds
\bea\label{oisa}
N=\frac{ m}{g^2}\ \ \frac{\sqrt{({\bar c}^2-1)(c^2-1)}}{(1-\kappa^2)(c+{\bar c})}\ \ (h+{\bar h})\ ,\ \ \ \ \ \ \ \ \ 
-\infty<N<\infty\ .
\eea

The  target space background without  torsion deserves a special mention. 
In  this  case $h={\bar h}=0$, 
the metric \eqref{isussossais} reduces to
the one from  Ref.\cite{Fateev:1996ea}. 
V.A. Fateev   used the coordinates $z$ and $(\chi_1,\chi_2)$ 
related to  $(u,v)$ through   the formula 
\eqref{sosssiuaus} with  $R=1$ and compactified as in \eqref{usooiasi}.\footnote{
The five parameters $(u,a,b,c,d)$ from  Ref.\cite{Fateev:1996ea}, 
 subject to  the  constraints
$(u+d)^2=a^2+c^2,\ d^2=b^2+c^2$, are related to the set   $(g,\,p ,\,\kappa)$  through   the formulas
$$
\frac{a}{u}=\frac{1}{m}\ ,\ \ \ \ \ 
\frac{b}{u}=\frac{\kappa}{m}\ ,
\ \ \ \ \ 
\frac{c}{u}=\frac{\kappa}{2m^2}\  (p^2-p^{-2})\ ,\ \ \ \ \ 
\frac{d}{u}=- \frac{\kappa}{2m^2}\ (2\kappa+p^2+p^{-2})\ ,\ \ \ \ \ 
u=\frac{g^2}{4}\ ,
$$
where $m=+\sqrt{1+\kappa^2+\kappa p^2+\kappa p^{-2}}$.}
The ZCR for  the Fateev model  was not known before.
It  is merely a specialization of the general
case for $\nu=\sigma=0$.  

The   three-dimensional target space background\ \eqref{isussossais}
can be used to  
build an  integrable NLSM with $D=2$.
Let us  set $h={\bar h}=0$,  $g^2=\kappa\ {\tilde g}^2\ p^2$  and consider the   limit $p\to\infty$ with
$({\tilde  g},\, \kappa)$  held constant.  
This  formal procedure  yields 
\bea\label{sossaas}
G^{(2)}_{\mu\nu}\ \rd X^\mu\rd X^\nu=
\frac{1}{{\tilde  g}^2}\ \bigg[\, \frac{(\rd z)^2}{(1-z^2)(1-\kappa^2 z^2)}\ 
+ \frac{ (1-z^2)\,  (\rd v)^2}{ 1-\kappa^2\,z^2}\, \bigg]\ .
\eea
The coordinate $w$ does not appear at this limit, consequently  Eq.\eqref{sossaas} can be interpreted
as a metric for some  NLSM with two-dimensional target space. 
In fact, 
this metric  is  equivalent to the sausage  metric
from Ref.\cite{Fateev:1992tk}, provided $v\sim v+2\pi$.
One can check that the equations of motion for this NLSM 
admit the ZCR which follow from the general ZCR
as $\eta\to -\ri\pi  \tau$  (see Eq.\eqref{ksisusossai}). 
Taking the 
limit,
Eqs.\eqref{uissasaus}-\eqref{issusa} with  $0<u<\pi$ and $\nu=\sigma=0$  yield
\bea\label{ystsossai}
e_v= {\tilde g}^{-1}\ \vartheta^2_3(0,q^2)\ ,\ \ \ \ \  \ \ e^{\pm}_v= {\tilde g}^{-1}\ 
\re^{\pm \ri (\phi_+-u)}\ 
\frac{  \vartheta_1(u, q^2) \vartheta_3(0, q^2)}{  \vartheta_4(u, q^2)\vartheta_2(0, q^2)} \ , \ \ \  \ \ 
e_w^{\pm}=0\ ,
\eea
whereas  $\phi_+=\phi_-$.
Then   Eqs.\eqref{sopsosaosa}-\eqref{tsososa}  with $\eta= -\ri\pi  \tau$ can be applied literally.

\section{  Ricci  flow with torsion}

\subsection{One-loop renormalization}

We turn now to a discussion of the renormalization  effects in the NLSM under consideration.
The  RG flow equations for a general target space background were computed up to two loops in Refs.\cite{Honerkamp:1971sh,
Polyakov:1975rr, Friedan:1980jf, Witten:1983ar,  Braaten:1985is, Fradkin:1985ys,  Metsaev:1987bc}.
At leading order, the  equations 
can be written  in   somewhat  symbolic form\ \eqref{sosasaias} \cite{Braaten:1985is}.  For  practical purposes,
it is useful to rewrite  them  in terms of the symmetric  Ricci tensor $R_{\mu\nu}$ associated with
the  Levi-Civita connection  $\nabla_\mu$:
\bea\label{siisisa}
{\dot G}_{\mu\nu}&=&-\ \big(\, R_{\mu\nu}-{\textstyle\frac{1}{4}}\ {H_{\mu}}^{\sigma\rho}\,H_{\sigma\rho\nu} 
+\nabla_\mu V_\nu+ \nabla_\nu V_\mu\,\big)\nonumber\\
{\dot B}_{\mu\nu }&=&-
\big(\, {\textstyle\frac{1}{2}}\ \nabla_\sigma {H^{\sigma}}_{\mu\nu}-V_\sigma\, {H^{\sigma}}_{\mu\nu }\,\big)\ .
\eea
Here the dot stands for the (partial) derivative with respect to the RG ``time'' which is proportional to the logarithm of
the   RG energy scale $E$:
\bea\label{sooas}
t=-\frac{1}{2\pi}\ \log\Big(\frac{E}{E_*}\Big)\ ,
\eea
where   $E_*$ (the integration constant of the RG flow equations) sets the ``physical scale'' for the NLSM.
Some clarification  is needed for  the terms  depending on    an arbitrary one-form $V_\mu$.
The general form of an  infinitesimal   RG transformation should
admit  the possibility of
various coordinate  transformations \cite{Friedan:1980jf}.
Under an   arbitrary   infinitesimal reparameterization
$\delta G_{\mu\nu}=-(\nabla_\mu V_\nu+ \nabla_\nu V_\mu)\,\delta t$, 
whereas 
$\delta B_{\mu\nu}=V_\sigma\, {H^{\sigma}}_{\mu\nu}\,\delta t+ {\tilde \delta}B_{\mu\nu}$
with ${\tilde \delta}B_{\mu\nu}=-(\partial_\mu V_\nu- \partial_\nu V_\mu)\,\delta t$.
The variation ${\tilde \delta} B_{\mu\nu}$ is a  pure  gauge transformation which
does not affect the  torsion strength and therefore  can be neglected.
Thus the terms with $V_\mu$ incorporate the  effects of all possible diffeomorphisms
and can be chosen arbitrarily  (to a certain extent, see subsection \ref{dilaton} bellow)
in order to simplify the equations.
In what follows we assume that
there exists  a   diffeomorphism  generating
function    $\Psi$ such that
\bea\label{soisisa}
V_\mu= \partial_\mu\Psi\ .
\eea

To establish  the one-loop renormalizability 
of the    finite-parameter family   of NLSM,
it is sufficient to demonstrate that, for
some choice of the   diffeomorphism  generating function,
the RG flow equations  can be   satisfied  
by allowing    the parameters $(p,h,{\bar h},\kappa)$
and the string tension $g$ to be $t$-dependent.
With a brief look at   \eqref{siisisa} we conclude 
that $\Psi$  is
transformed as a scalar  under $t$-independent coordinate transformations only.
Therefore it  essentially depends
on a 
 choice of the
coordinate system or, widely speaking,  on the    RG scheme.
It turns out that  the $(u,v,w)$-coordinate frame  is very useful 
for adjusting  the
diffeomorphism  generating function.
Using these coordinates
one can show that Eqs.\eqref{siisisa},\,\eqref{soisisa} are indeed satisfied, provided
\bea\label{psososa}
\re^{2\Psi}=\re^{2\Psi_0}\ \rho(u)\rho(-u)\ ,
\eea
where $\rho(u)$ is given by \eqref{kksisisa} and $\Psi_0$  is an arbitrary
coordinate-independent constant. More precisely, with this  choice of $\Psi$ and
for $G_{\mu\nu},\ B_{\mu\nu}$  defined by Eqs.\eqref{uissasaus}-\eqref{sosoasa},
the  one-loop RG flow equations    are  reduced to the following closed system
of ODE: 
\bea\label{sosasias}
&&{\dot \kappa}=-\frac{g^2}{m^2}\ \kappa\, (1-\kappa^2)\nonumber\\
&&{\dot g} =\frac{g^3}{4 m^4}\ (1-\kappa^2)^2\ \big(\,
1-N^2\, g^4\, \big)\nonumber\\
&&{\dot p}=0\\
&&{\dot c}=\frac{g^2}{ m^2}\ \frac{(c^2-1)(\,\kappa^2\, c{\bar c}+1)}{(c+{\bar c})}\nonumber\\
&&{\dot {\bar c}}=\frac{g^2}{ m^2}\ \frac{({\bar c}^2-1)(\,\kappa^2\, c{\bar c}+1)}{(c+{\bar c})}\ ,\nonumber
\eea
where $(m,c,{\bar c},N)$  are expressed in terms of
the 
independent set
$(p,h,{\bar h},\kappa)$  as in Eqs.\eqref{psooisisaais},\,\eqref{oisa}. 

The solution of the system \eqref{sosasias}  is a  rather straightforward exercise.
First of all,   it is evident that $P$\,\eqref{paosaos} and 
$R$\,\eqref{sosisa} are   the   first integrals.
Then we should recall  that
the path-integral  quantization procedure 
requires that the ${\cal H}$-flux, 
$ \frac{1}{\pi} \int_{{\cal M}_3}{\cal  H}$, must be  an integer.  
Thus, $N$ given by Eq.\eqref{oisa}   must  be the  first integral as well as $P$ and $R$.
This, of course, can be easily tested.
Note that the  flux essentially depends on a  choice of
compactification of the Killing coordinates, and therefore  the   RG invariance 
of \eqref{oisa}  provides an additional  support  for the assumptions \eqref{usooiasi}.
A further analysis of the first two equations  in \eqref{sosasias}
 yields
one more first integral   which can be
chosen in the form
\bea\label{jsaisaoisa}
M^2=\frac{m^2}{4\kappa g^4}-\frac{(1+\kappa)^2}{4\kappa}\ N^2\ .
\eea
To summarize, for  $N\not=0$ the system of   ODE\ \eqref{sosasias} possesses the following
complete set of the first integrals: 
\bea\label{sososaisa}
N\not=0\ \ \ :\ \ \ \ \ \ \  (P, R, M, N)\ .
\eea
The condition $N=0$  implies  $R^2=1$. In this case 
the complete set of  the first integrals  can be chosen as follows
\bea\label{aospaso}
N=0\ \ \ :\ \ \ \ \ \ \  (P, M, L)\ \ \ \ \ {\rm with}\ \ \ \
L=\frac{h}{\sqrt{\kappa}}=-\frac{{\bar h}}{\sqrt{\kappa}}\ .
\eea
The first  integral  $L^2$ can be also  naturally  introduced for  nonvanishing $N$.
Indeed, as it follows from the first and  the last two equations in \eqref{sosasias} 
\bea\label{saossai}
c\,{\bar c}=\frac{1+L^2\kappa}{\kappa (L^2+\kappa)}\ ,\ \ \ \ \  \ c-{\bar c}=\frac{R^2-1}{R^2+1}\ \frac{1-\kappa^2}{\kappa (L^2+\kappa)}\ ,
\eea
where  $L^2$ stands  for  the RG invariant  which can be alternatively defined by  Eq.\eqref{aospaso}  in  the case $N=0$.
The first  integrals   $L^2$,    $R$ and the ratio $\frac{N}{M}$ are functionally dependent
(see Eqs.\eqref{ssosa},\,\eqref{asosasi} bellow).

The existence of  a complete set of  first integrals makes it possible to 
integrate
the ODE system \eqref{sosasias} explicitly.
One just needs to  substitute $\kappa$,  which is  an  elliptic modulus associated with
the elliptic nome $q^2$,    for  an    elliptic modulus $k$ related to the elliptic nome $q$, or 
equivalently,
to
perform Landen's transformation $\kappa\to k=\frac{2\,\sqrt{\kappa}}{1+\kappa}$.
Then  a simple calculation yields the result 
\bea\label{ossshai}
\re^{ t}=\bigg|\frac{g^2-g_0^2}{g^2+g_0^2}\bigg|^{\frac{1}{ b^2}}\
\bigg|\frac{1+N\, g^2}
{1-N\, g^2}\bigg|^{N}\
\bigg|\frac{ M\, g^2+ P }
{M\, g^2- P}\bigg|^{MP}\ ,
\eea
and
\bea\label{saopspsao}
\kappa=\frac{1-\sqrt{1-k^2}}{1+\sqrt{1-k^2}}\  ,\ \ \ \ \ k^2=\frac{(1+N\, g^2)(1-N\, g^2)}
{(M\, g^2+ P)(M\, g^2- P)}\ ,
\eea
where
\bea\label{sioiosa}
b^2=+\frac{1}{\sqrt{(N^2+M^2)(1+P^2)}}\ ,\ \ \ \ \ g_0^2=+ \sqrt{\frac{1+P^2}{N^2+M^2}}\ .
\eea
To evaluate  the running coupling constant $g=g(t)$  as a function
of the RG scale $E$\ \eqref{sooas} requires an inversion of the  relation \eqref{ossshai}.
Note that, as follows from \eqref{saopspsao},
$\big|\frac{ M\, g^2+ P }
{M\, g^2- P}\big|\not=0,\infty$ as $0\leq \kappa\leq 1$. 
By following the logical structure of quantum field theory,
the   renormalized parameters of the  target space  background should be expressed in terms of
the running coupling constant and the RG invariants. Eqs.\eqref{saopspsao}
allow one to do this for  $\kappa$.  The  relations \eqref{saossai} define $c,\,{\bar c}>1$  unambiguously
through the solution of a quadratic equation. Then, using Eqs.\eqref{psooisisaais} one can determine $(h^2,\,{\bar h}^2)$.
The signs of $h$ and ${\bar h}$ can recovered from \eqref{oisa}. Note  that $\sgn(h+{\bar h})=\sgn(N)$.
Finally, the  parameter $p$ is a RG invariant itself.

Finishing with the solution of Eq.\eqref{siisisa}, let us   note that the   constant $\Psi_0$ in Eq.\eqref{psososa} does not
contribute to the RG flow equations. However, if we set it to be
\bea\label{asoisais}
\exp\big(2\Psi_0(t)\big)=\sqrt{ \frac{g\,\kappa}{(1-N^2 g^4)(1-\kappa^2)}}\ ,
\eea
then 
the  following  relation  is  satisfied
\bea\label{sisai}
\partial_t\big(\, 2 \Psi-\log\sqrt{G}\ \big)=-{\textstyle\frac{1}{4}}\ \big( -G^{\mu\nu}\, { R}_{\mu\nu}+
{\textstyle \frac{1}{12}}\ H_{\mu\nu\lambda}\,H^{\mu\nu\lambda}+
4 { \nabla}_\mu\Psi\,{ \nabla}^\mu\Psi-4\nabla_\mu\nabla^\mu\Psi\,\big)\ ,
\eea
and the  constructed  set of fields   $(\,G_{\mu\nu}$, $B_{\mu\nu},\Psi\,)$ provides  a solution to
the coupled system of   PDE \eqref{siisisa},\eqref{soisisa},\,\eqref{sisai}.
Note that, in practical calculations, it is usually desirable to 
rewrite  \eqref{psososa}  in terms of the  doubly periodic function $z=z(u,q)$
defined in \eqref{sossaius}:
\bea\label{sossai}
\exp\big(2\Psi(u,t)\,\big)=\exp\big(2\Psi_0(t)\big)\ \  \frac{\vartheta^2_4(0,q^2)}{\vartheta^2_4(u,q^2)}\ \ 
\frac{(c+1)({\bar c}-1) (1-\kappa^2 z^2)}{(c+z)({\bar c}-z)(1-\kappa^2)}\ .
\eea
Here
the elliptic parameter $q$  should be treated as a function of the RG time  and, as  it follows from 
the first   equation in \eqref{sosasias},
\bea\label{saosaias}
\frac{\dot q}{q}=-\frac{g^2}{\vartheta^4_3(0,q^2)\,m^2}=-\frac{1}{G_{uu}}\ .
\eea

\subsection{\label{UV} Ultraviolet behavior}

As $\kappa=1$ 
the ODE system\ \eqref{sosasias} possesses  constant solutions  $(p, h, {\bar h}, g)=(p_0, h_0, {\bar h}_0, g_0)$
which can be specified by the set of numbers
\bea\label{oppsasp}
\Big(\, \alpha,\beta,\delta,\, b\,\big|\, -\frac{\pi}{2}< \alpha,\, \beta<\frac{\pi}{2},\ \ \
0< \delta<\pi\, ,\ b>0\, \Big)
\eea
through the  relations
\bea\label{osaosa}
h_0=\tan(\alpha)\ ,\ \ \  \ {\bar h}_0=\tan(\beta)\ ,\ \ \  
p_0=\cot\Big(\frac{\delta}{2}\Big)\ ,\ \ \ \ 
g_0^2=
\frac{ b^2}{\sin^2(\delta)}\ .
\eea
Let  ${\cal S}^{(\alpha,\beta)}_{(\delta| b)}$ be
the RG trajectory which asymptotically approaches 
the  constant solution characterized by  a  given set \eqref{oppsasp} as $t\to-\infty$. 
The values of the  functionally independent   RG invariants   for  ${\cal S}^{(\alpha,\beta)}_{(\delta| b)}$
are determined through the formulas
\bea\label{ssosa}
P=\cot(\delta)\ ,\ R=\frac{\cos(\alpha)}{\cos(\beta)}\ ,\ \ \ 
N=\frac{1}{b^2}\ 
\sin(\alpha+\beta)\sin(\delta)\ ,\ \ \ M=\frac{1}{b^2}\ \cos(\alpha+\beta)\sin(\delta)\ ,
\eea
whereas the RG invariant $L^2$ from Eq.\eqref{saossai} is given by
\bea\label{asosasi}
L^2=\frac{\sin^2(\alpha)+\sin^2(\beta)}{\cos^2(\alpha)+\cos^2(\beta)}\ .
\eea  

It is interesting to look at  the  asymptotic form of 
the target space background 
corresponding to the RG trajectory ${\cal S}^{(\alpha,\beta)}_{\delta |b}$ at  large negative
$t$.
For this purpose, let us  cut the chart defined by  \eqref{isosasai} into  the  three pieces $U^{(0)}$, $U^{(1)}$ and 
$U^{(2)}$
depending on  the value of the  coordinate $z$:
$-1+\epsilon\leq z\leq 1-\epsilon$, $0<1-z\leq \epsilon$ and  $0<1+z\leq \epsilon$, respectively.
Here $\epsilon$ stands for  some  small number  which is, in the case $1-\kappa\ll 1$,  can be chosen to satisfy
both 
conditions  $\epsilon\ll 1$ and  $\epsilon\gg 1-\kappa$ simultaneously. 
Then on the chart $U^{(0)}$ covering  the  central region  of ${\cal M}_3$,
we replace $z$ by $\rho$:
\bea\label{asossai}
\rho=\frac{2K}{\pi}\ \Big(u-\frac{\pi}{2}\Big)\ ,
\eea
where  $u$ stands  for our  original variable\ \eqref{sossaius} and $K$ is    the elliptic 
quarter-period
associated with the nome $q^2$, i.e., $K=\frac{\pi}{2}\ \vartheta^2_3(0,q^2)\approx
\frac{1}{2}\, \log(\frac{8}{1-\kappa})$.  
It is  straightforward to see  that  the metric on the chart $U^{(0)}$ is  approximated by the  form
\bea\label{ytsosissa}
G^{(UV)}_{\mu\nu}\,\rd X^\mu\rd X^\nu\big|_{U^{(0)}}\approx\frac{4}{ b^2}\ \Big[\, (\rd \rho )^2+
\big(\rd \chi^{(\alpha)}_1\big)^2+
\big(\rd \chi^{(\beta)}_2\big)^2+2\,  \cos(\delta)\, \rd \chi^{(\alpha)}_1\rd \chi^{(\beta)}_2\,  \Big],
\eea
whereas the torsion strength ${\rm H}\approx 0$.
Here we use  $\chi_1^{(\alpha)}=\cos(\alpha)\, \chi_1$,
$\chi_2^{(\beta)}=\cos(\beta)\, \chi_2$, and as  it follows from Eqs.\eqref{usooiasi},
\bea\label{wesossai}
 \chi^{(\alpha)}_1\sim \chi^{(\alpha)}_1+2\pi\,\cos(\alpha)\, ,\ \ \ \ \chi^{(\beta)}_2\sim\chi^{(\beta)}_2+
2\pi\, \cos(\beta)\, .
\eea
Thus, in the  central region,  the metric  is almost flat and
the  target manifold  ${\cal M}_3$    is well  approximated
by the Cartesian  product
of  the  two-torus  and the  line segment of  total length $\ell\approx 2b^{-1}\, \log(\frac{2}{\epsilon})$.

Similarly to\ \eqref{asossai}, at  the charts  $U^{(1)}$ and  $U^{(2)}$ we replace   the coordinate $z$ by 
\bea\label{soissiusa}
\rho_1=\frac{2K}{\pi}\ u\ ,\ \ \ \ \ \ \ \ \rho_2=\frac{2K}{\pi}\ (u-\pi)\ ,
\eea
respectively. 
The target space background in the region 
covered by the chart $U^{(1)}$ is approximated as follows:
\bea\label{ososapsa}
G^{(UV)}_{\mu\nu}\,\rd X^\mu\rd X^\nu\big|_{U^{(1)}}&\approx&
\frac{4}{ b^2 }\ \bigg[\,\big(\rd \rho_1 \big)^2+
\frac{ \cos^2(\beta)\, \sinh^2(\rho_1)}{\cosh(\rho_1-\ri\beta){\cosh(\rho_1+\ri\beta)}}\ 
 (\rd\psi_1)^2 \nonumber\\
&+&
\frac{\cosh^2 (\rho_1)}
{\cosh(\rho_1-\ri\beta){\cosh(\rho_1+\ri\beta)}}\   \big(\rd \chi^{(\alpha,\delta)}_1  \big)^2
\, \bigg]\\
\frac{1}{2}\ B^{(UV)}_{\mu\nu}\,\rd X^\mu\wedge \rd X^\nu\big|_{U^{(1)}}
&\approx&-\frac{4}{b^2}\ \frac{\sin(\beta)\ \sinh^2(\rho_1)}
{\cosh(\rho_1-\ri\beta){\cosh(\rho_1+\ri\beta)}}\ 
\  \rd \chi^{(\alpha,\delta)}_1\wedge \rd\psi_1\ .\nonumber
\eea
Here we use the notations
\bea\label{oisasosa}
\psi_1=\chi_2+\frac{\cos(\alpha)\cos(\delta)}{\cos(\beta)}\ \  \chi_1\ ,\ \ 
\ \ \ \ \ \ \chi_1^{(\alpha,\delta)}=\sin(\delta)\, \cos(\alpha)\, \chi_1\ .
\eea
There is no need to present similar formulas for the region
covered by the chart $U^{(2)}$. They are
obtained by substituting
$\chi_{1}\leftrightarrow\chi_{2},\ 
\psi_{1}\leftrightarrow\psi_{2}$ and
$\alpha\leftrightarrow\beta$ in the above expressions.
Since  $\rho=\rho_1-K$,
the  metrics \eqref{ytsosissa} and   \eqref{ososapsa}
are smoothly sewed together
as  $(-\rho)\sim \frac{1}{2}\ \log(\frac{2}{\epsilon})\gg 1$
and
$\rho_1\sim\frac{1}{2}\,
\log(\frac{4\epsilon}{1-\kappa})\gg 1$. In this domain the 
torsion strength corresponding to the torsion potential\ \eqref{ososapsa} becomes  of the order
$\frac{1-\kappa}{\epsilon}\ll 1$.

Of course, the  metric  in the r.h.s. of \eqref{ytsosissa}  combined with    ${\rm H}=0$,  
provides  a stationary solution of the RG flow equations.
The background \eqref{ososapsa} can be made  into
an RG fixed point in the precise sense of the world by an appropriate definition of
the RG transformation or, in stringy speak, to introducing the dilaton field
\bea\label{opaosp}
\re^{2\Phi}=\frac{\cos^2(\beta)}{\cosh(\rho_1-\ri\beta)\,\cosh(\rho_1+\ri\beta)}\ .
\eea
Then, the limiting form of
$( G^{UV}_{\mu\nu}\,,B^{(UV)}_{\mu\nu})_{U^{(1)}}$ as $t\to -\infty$,
together with  $\Phi$ satisfy the so-called string equations \cite{ Fradkin:1985ys, Callan:1985ia} (the conditions for Weyl
invariance to hold  in the NLSM in the lowest nontrivial approximation):\footnote{
This  solution of the string equations  is well known.
Without regard to compactification conditions\ \eqref{usooiasi},\,\eqref{oisasosa}, it coincides with
the marginal deformation  of  the  Euclidean version of  $SL(2,{\mathbb R})$ (i.e., ${\mathbb H}_3^+)$
 WZWN background   
(see e.g.\cite{Forste:1994wp} and references therein). 
The  symmetric $ {\mathbb H}_3^+$-background occurs in the properly taken limit  $\ri\,\beta,\,\ri\,\alpha\to \infty$. 
A compact version of the  background
(see Eqs.\eqref{sosussysais},\,\eqref{opsosa} below) 
were originally introduced  in Refs.\cite{Hassan:1992gi,Kiritsis:1993ju}.}
\bea\label{xsiisisa}
&&R_{\mu\nu}-{\textstyle\frac{1}{4}}\ {H_{\mu}}^{\sigma\rho}\,H_{\sigma\rho\nu}
+2\nabla_\mu \nabla_\nu\Phi=0\nonumber\\
&&{\textstyle\frac{1}{2}}\ \nabla_\sigma {H^{\sigma}}_{\mu\nu}-\nabla_\sigma\Phi\, {H^{\sigma}}_{\mu\nu }=0
\\
&& -G^{\mu\nu}\, { R}_{\mu\nu}+
{\textstyle \frac{1}{12}}\ H_{\mu\nu\lambda}\,H^{\mu\nu\lambda}+
4 { \nabla}_\mu\Phi\,{ \nabla}^\mu\Phi-4\nabla_\mu\nabla^\mu\Phi=const\nonumber\ .
\eea
Note that the dilaton \eqref{opaosp} does not  vanishes  as $\rho_1\gg 1$
but   approaches to $ const -\rho_1$. Therefore,   the constant in the r.h.s. of the last equation in \eqref{xsiisisa}
is equal to $ b^2$.
It occurs because 
the region  of  ${\cal M}_3$ in the vicinity $|z-1|=\epsilon$ remains fixed
with respect to the coordinate frame at $U^{(0)}$. However, since $\rho=\rho_1-K(t)$,
it  ``flows''  uniformly  without changing its shape    with respect to the 
chosen coordinate frame  at   $U^{(1)}$.

Of course, the string equations is   a  stationary version of\ \eqref{siisisa},\,\eqref{soisisa},\,\eqref{sisai}
and
\bea\label{ososa}
\Phi=\lim_{t\to-\infty\atop
\rho_1-{\rm fixed}}\big(\Psi-\Psi_0\big)\ .
\eea
It should be emphasized that, contrary to  the  diffeomorphism  generating function $\Psi$,
the dilaton  scalar field   is  a   RG scheme-independent (universal), characteristic
of the {\it critical} target space background\ \cite{Fradkin:1985ys}.


\subsection{Infrared behavior}
 
Let us  consider  first the case with    $N\not=0$  or, equivalently,  $\alpha+\beta\not= 0$. Then the
RG trajectory ${\cal S}^{(\alpha,\beta)}_{(\delta| b)}$ 
can be
extended to a complete {\it eternal} solution, i.e., it is defined for  $-\infty<t< +\infty$.
As it follows from Eqs.\eqref{ossshai},\,\eqref{saopspsao} 
the parameter $\kappa$ becomes   zero at the limit   $t\to+\infty$ whereas
\bea\label{saioosas}
\lim_{t\to +\infty}\, {g^2}=\frac{1}{|N|}\ .
\eea
It is also straightforward to see  that there exist the limits
\bea\label{sososao}
h_*=\lim_{t\to +\infty} h\ ,\ \ \ \ \ {\bar h}_*=\lim_{t\to +\infty} {\bar h}\ ,
\eea
and their values depend  on the RG invariant $R$ and on the sign of $N$ only:
\bea\label{ospaso}
&{\bar h}_*=0\ ,\ \ \ \ &h_*=\sgn(N)\ \ \frac{1}{2}\ (R^{-1}-R)\ \ \ \ \ \ \ \ \ \ \ \ (0<R\leq 1)\nonumber\\
&{ h}_*=0\ ,\ \ \ \  &{\bar h}_*= \sgn(N)\ \ \frac{1}{2}\ (R-R^{-1})\ \ \ \ \  
\ \ \ \ \ \ \
(R\geq 1)\ .
\eea
To describe  the target space backgrounds corresponding to  the infrared fixed point
of the RG flow, it is convenient to use
the RG invariants $R,\,N$ and the 
Hopf coordinates 
$(\theta,\chi_1, \chi_2)$ with   $z=\cos(2\theta) $ and $\chi_a$   defined by\ \eqref{sosssiuaus}.
Then for any $R>0$ one founds
\bea\label{sosussysais}
&&G^{(IR)}_{\mu\nu}\, \rd X^\mu \rd X^\nu=4\,|N|\ \bigg[\, (\rd\theta)^2+
\frac{R^2\cos^2(\theta)\ (\rd\chi_1)^2}{\cos^2(\theta)+R^2\sin^2(\theta)}+
\frac{\sin^2(\theta)\ (\rd\chi_2)^2}{\cos^2(\theta)+R^2\sin^2(\theta)} \, \bigg]\nonumber\\
&&\frac{1}{2!}\ 
B^{(IR)}_{\mu\nu}\, \rd X^\mu\wedge\rd X^\nu=-\frac{4\,N\ R^2\sin^2(\theta)}{\cos^2(\theta)+R^2\sin^2(\theta)}\ \ 
\rd\chi_1\wedge\rd\chi_2\ ,
\eea
and for the dilaton field
\bea\label{opsosa}
\exp\big(2\Phi^{(IR)}\,\big)=  \frac{1}{\cos^2(\theta)+R^2\, \sin^2(\theta)}\ .
\eea
As it was already mentioned,   the RG invariant $N$  must satisfy the   quantization condition
\bea\label{ksoosai}
|N|=\frac{n}{16\pi}\ \ \  \ \ \ \ {\rm with}\ \ \ \ \ n=1,2,\ldots\ .
\eea
The set of fields $\big(G^{(IR)}_{\mu\nu},\,B^{(IR)}_{\mu\nu},\,\Phi^{(IR)}\big)$ 
obeys  Eqs.\eqref{xsiisisa}. This  solution
of the string equations was introduced in Ref.\cite{Hassan:1992gi,Kiritsis:1993ju} and it is
usually referred as to  marginally deformed WZWN  model.

Let us turn now to the case  $N=0$.
The  RG trajectory  ${\cal S}^{(\alpha,-\alpha)}_{(\delta| b)}$ corresponds to the  ancient solution
terminating at   $t=0$ when  the running  coupling constant becomes infinite  (see Eq.\eqref{ossshai}). 
As $t\to -0$,  the torsion strength vanishes  whereas
the metric asymptotically approaches to the
round sphere metric\ \eqref{ussaias} whose   radius  $\frac{2}{g(t)}$  shrinks to zero  at $t=0$\ \cite{Polyakov:1975rr}:
\bea\label{oaspas}
\frac{4}{g^2}\sim -2\, t=\frac{1}{\pi}\ \log\Big(\frac{E}{E^*}\Big)\ .
\eea

\subsection{\label{dilaton} Comment on  the diffeomorphism generating function}

We finally discuss the relevance of the diffeomorphism dependent terms ($V$-terms bellow) in the Ricci flow equations.
Since the  diffeomorphism generating function 
depends on a choice of the coordinates,  
we  can use it to  simplify 
the  general form  of the Ricci flow equations  somewhat. Namely,   it seems natural   to exclude the  $V$-terms  from \eqref{siisisa} 
by a proper choice of 
the  coordinate  system $Z^{\mu}$,  ``moving'' 
with respect to the frame $X^{\mu}=(u, v,w)$.  The desirable coordinate 
frame is defined by the equation 
\bea\label{issassua}
\frac{\rd  Z^{\mu}}{\rd t}:=\Big(\frac{\partial Z^\mu}{\partial t}\Big)_X+ G^{\mu\nu}\ \partial_\nu\Psi=0\ .
\eea
Let us chose the 
new coordinates  in the form    $Z^\mu=(Z, v, w)$  with   $Z=Z(u,q)$ and  $q=q(t)$. Then,  Eq.\eqref{issassua}
combined with   \eqref{saosaias}, yields
\bea\label{sosia}
q\, \partial_q Z-\partial_u \Psi\ \partial_u Z=0\ .
\eea
This linear PDE  should be supplemented by the    initial condition
$Z|_{q=q_0}=Z_0(u)$.
Of course,   the desirable coordinate system 
is defined up to $t$-independent diffeomorphisms, so that  $Z_0(u)$ is  a rather arbitrary monotonic function of $u\in (0,\pi)$.

To be  more specific at this point, let us consider  the RG trajectory ${\cal S}^{(\alpha,-\alpha)}_{(\delta| b)}$.
Then, as it follows from\ \eqref{sossai},
\bea\label{saooasi}
\partial_u \Psi=-\frac{1}{2}\ \partial_u\log\big(\,\vartheta_4^2(u,q^2)+L^2\, \vartheta_1^2(u,q^2)\, \big)\ ,
\eea 
where $L=\tan(\alpha)$ stands for  the RG invariant\ \eqref{aospaso}.  The initial condition  can be taken at $q=0$ with
$Z_0(u)=\cos(u)$. Given  this initial setup,
the  solution of \eqref{issassua} is constructed as a power series in $q$:
\bea\label{wsossaiss}
Z=\cos(u)+2L^2\, \sin(u)\,\sin(2u)\, q+ 2 \, \sin(u)\,\sin(2u) \, (1-4\,L^4\ \cos^2(u)\,\big)\ q^2+
O( q^{3})\ .
\eea
Note that the $n$-th term  of this  series is a polynomial in  $L^2$  of order $n$.
The  partial summation of the series yields
\bea\label{isauasuy}
Z=z+\frac{L^2}{2}\, (1-z^2)\  \log\Big(\frac{1+ \kappa\, z}{1-\kappa\, z}\Big)+O(L^4)\ ,
\eea
where $z=z(u,q)$ and $\kappa=\kappa(q)$ are  given by\ \eqref{sossaius} and  \eqref{ystsosoasi}, respectively.
Eq.\eqref{isauasuy} implies  that, 
 in the torsion-free case,   the metric\ \eqref{isussossais}
with $h={\bar h}=0$   satisfies
${\dot G}_{\mu\nu}=-R_{\mu\nu}$.
In fact,  it  was discovered   by V.A.  Fateev as a  brute-force solution to  this Ricci flow equation.
The expansion  \eqref{isauasuy} also  suggests  to consider  $(z,\kappa)$ as an independent set of variables replacing the variables $(u,q)$.
It is then straightforward to check that
\bea\label{rsossai}
Z=Z[z,\kappa]\ :\ \ \partial_\kappa Z-\frac{L^2\, z\, (1-z^2)}{1+\kappa L^2-\kappa(\kappa+L^2)\, z^2}\   \partial_z Z=0\ ,\ \ \ \ Z[z,0]=z\ .
\eea
The solution of this  Cauchy problem   can be obtained  by the  method of characteristic:
\bea\label{rsususa}
Z=\frac{(1+z)(1+\kappa z)^{L^2}-(1-z)(1-\kappa z)^{L^2}}{(1+z)(1+\kappa z)^{L^2}+(1-z)(1-\kappa z)^{L^2}}\ .
\eea
It is  a nonsingular monotonic  function of $z\in[\,-1,\,1\,]$    for any     $0\leq \kappa<1$.
However, as $\kappa\to 1$,  the branch points  at $z=\pm\kappa^{-1}$ approach   the ends of  the  segment. For this reason
the regions    $(z,\kappa\,|\, |1-\kappa|\ll 1,\     |z\pm 1|\ll  1\,)$ need a special attention.
As it has been  discussed in subsection\ \ref{UV}, the target space backgrounds  in these domains are  
asymptotically approaching  the  solutions of the string equations\ \eqref{xsiisisa}, and
the each  critical  background required
the dilaton field
which cannot  be  absorbed by a nonsingular coordinate transformation.

In the case $N\not=0$  we  can still chose   $(z,\kappa)$ as an independent set of variables. Then,
using Eqs.\eqref{saossai},\,\eqref{sossai}, the linear 
PDE \eqref{sosia} can be brought to the   form 
\bea\label{xrsossai}
\partial_\kappa Z+F(z,\kappa)\ \partial_z Z=0\ ,
\eea
where
\bea\label{sisaiusa}
F(z,\kappa)=-\frac{ \big(\,2\kappa\,L^2\,z+A\, (1+\kappa^2 z^2)\,\big)\, (1-z^2)}{2\kappa\,
\big(\,1+\kappa\, L^2-A\,(1-\kappa^2)\, z-\kappa(\kappa+L^2) \,z^2\,\big)}
\eea
and   $A$ stands for the RG invariant
\bea\label{osaasia}
A=\frac{R^2-1}{R^2+1}\ .
\eea
For $A\not=0$,
the solution  of the characteristic curve equation 
\bea\label{saosiosa}
\frac{\rd z}{\rd \kappa}=F(z,\kappa)\ ,
\eea
is not available in a closed  form; however, its small-$\kappa$ asymptotic   can be easily found. 
A simple calculation shows that the function $Z$ can be chosen in the form
\bea\label{akisisa}
Z\approx\frac{(1+z)\ \big(\frac{1-z}{\sqrt{\kappa}}\big)^A-(1-z)\ \big(\frac{1+z}{\sqrt{\kappa}}\big)^{-A}}
{(1+z)\ \big(\frac{1-z}{\sqrt{\kappa}}\big)^A+(1-z)\ \big(\frac{1+z}{\sqrt{\kappa}}\big)^{-A}}\ \ \ \ \ \ \ \ \ \ \ \ (\kappa\ll 1)\ .
\eea
It  makes explicit  non-analytic properties of the 
coordinate transformation  $z\to Z$ at the limit  $\kappa\to 0$.
As $\kappa=0$ the target space background arrives at  the infrared    fixed-point
which   requires the introduction of  the  dilaton field.

Returning to the general   one-loop  RG equations,  
our analysis illustrates  the   role 
of the $V$- terms in Eq.\eqref{siisisa}. Namely,  it suggests that,  by means of a nonsingular reparameterization
of the target manifold,
these terms  can be excluded  everywhere except   the RG  fixed-point regime.

\section{Conclusion}

In this paper   we  have   found  the zero-curvature representation for
a   four-parameter family of  the classical  NLSM. In the   context of a   hierarchy
of  classical integrable systems the  new family  can be viewed as 
a three-parameter deformation
of  the  $SU(2)$ WZWN   model.
Also it contains, as a two-parameter subfamily, the Fateev  sausage model. Thus the 
work resolves the long-standing question of  classical 
integrability of that model.

We  have   discussed  some aspects of  the  perturbative quantization.
It was 
demonstrated the renormalizability of the integrable  family of NLSM at   the  lowest perturbative order.
The RG equations  at  the one-loop  order   describe a Ricci 
flow with torsion.  Therefore, among the  results  of this paper is  an  interesting
set of ancient and eternal solutions of the Ricci flow.
In all likelihood these   solutions  
correspond to 
the  multi-parameter family  of integrable  quantum fields theories.
Currently a non-perturbative description  is available  for the case
of the Fateev model only. 

An analysis of this paper is   explicitly concerned with
a  relation between a particular  classical  integrable NLSM and 
explicit solutions of the  Ricci flow. It seems extremely
desirable to get a more general  understanding about this  remarkable  relation,
which  may provide new  analytical insights  in  searching for
physically interesting string backgrounds.

\section*{Acknowledgments}

I  am   grateful  to
V.V. Bazhanov, V.A. Fateev,  D.H. Friedan, G.W. Moore
and A.B. Zamolodchikov for interesting discussions.
\bigskip

\noindent This  research was supported in part by DOE grant
$\#$DE-FG02-96 ER 40949.

\section{\label{Appendix} Appendix}


Here we describe  a general solution to  Eqs.\eqref{ossasaus}. 

First of all  one  needs   explicit formulas for 
the Levi-Civita spin  connection. In the case at hand 
non-vanishing components of the connection  read as follows:
\bea\label{sossis}
{\boldsymbol \omega}_\mu=\sum_{a=3,\pm}\omega^a_\mu\,\sigma_a \ :\ \ 
\omega^a_\mu=
\begin{cases}
-{\textstyle \frac{1}{2}}\ \big(\, \Omega^{+-}+\Omega^{-+}\, \big)\ e^{3}_\mu\ \ \ \ \ 
 \ \ \ \ \ \ \ \  \ \ \ \ \ \, a=3 \ \ {\rm and}\ \ \mu=u\ ,
\\
\pm {\textstyle \frac{1}{2}}\   \big(\, \Omega^{+-}-\Omega^{-+}\, \big)\ e^{\pm }_\mu
+
\Omega^{\pm\pm}\ e^{\mp }_\mu
 \  \ \ \ \ 
a=\pm \ \ {\rm and}\ \  \mu=v,\,w
\end{cases}
\eea
where $\sigma_\pm=\frac{1}{2}\, (\sigma_1\mp\ri \sigma_2)$ and  $\Omega^{\alpha\beta}\ (\alpha,\beta=\pm)$ stands for
\bea\label{sosasisa}
\Omega^{\alpha\beta}=\frac{1}{2\sqrt{G_{uu}}}\ \ 
\frac{e^\alpha_w\, \partial_u e^{\beta}_v- e^\alpha_v\,\partial_u e^{\beta}_w}
{e^{-}_v e^+_w-e^{+}_v e^-_w}\ .
\eea
These formulas combined with the definitions\ \eqref{sopsosaosa},\,\eqref{sosapsaosap}
allow  one to rewrite Eqs.\eqref{ossasaus} in explicit form.  Namely, 
the equations for ${\boldsymbol \omega}_\mu=\frac{1}{2}\, ({\boldsymbol \omega}_\mu^++
{\boldsymbol \omega}^-_\mu)$ are given by
\bea\label{xosososa}
&&\Omega^{++}=-\frac{\ri }{2 }\  f_+(\pi-2\ri\,\eta)\  \re^{-\ri(\phi_++\phi_-)}\nonumber\\
&&\Omega^{--}=\frac{\ri }{2 }\  f_+(\pi+2\ri\,\eta)\ \re^{\ri (\phi_+ +\phi_-)}\\
&&\Omega^{+-}+\Omega^{-+}=-\ri\, f_3(\pi+2\ri\,\eta)+
\frac{\ri}{2\sqrt{G_{uu}}}\ \,\partial_u (\phi_++\phi_-)\nonumber\\
&&\Omega^{+-}-\Omega^{-+}=\frac{1}{2\ri}\ \big(
f_+(\pi-2\ri\,\nu)\ \re^{-\ri (\phi_+-\phi_-)}+
f_+(\pi+2\ri \,\nu)\,
\re^{\ri (\phi_+-\phi_-)}\,\big)\ ,\nonumber
\eea
whereas the corresponding equations
for the antisymmetric part ${\boldsymbol \omega}^+_{\mu}-{\boldsymbol \omega}^-_\mu=
-\frac{\ri}{2}\  {\rm H}\, {\boldsymbol \gamma}_\mu$ read as
\bea\label{sopssopsapsa}
{\rm H}&=&f_+(\pi-2\ri\,\nu)\  \re^{-\ri (\phi_+-\phi_-)}-
f_+(\pi+2\ri \,\nu)\  \re^{\ri (\phi_+-\phi_-)}\nonumber\\
{\rm H}&=&2\, f_3\big(\pi-2\ri\,\nu)+\frac{1}{\sqrt{G_{uu}}}\ \partial_u (\phi_+-\phi_-)\  .
\eea
Here we use the notations \eqref{sosoosa} and  $\phi_\pm$ stand for $u$-dependent  phases  from the matrixes  
${\boldsymbol U}_\pm=\exp\big({\frac{\ri}{2}\phi_\pm\sigma_3}\big)$.
Eqs.\eqref{sopssopsapsa} can be immediately integrated 
and their  general  (one-parameter family)   solution is given by
\bea\label{osososp}
&&\re^{\ri (\phi_+-\phi_-)}=
\frac{\vartheta_3(\frac{u-\ri\sigma+\ri\nu}{2},q)\,\vartheta_4(\frac{u+\ri \sigma+\ri\nu}{2}, q)}
{\vartheta_3(\frac{u+\ri\sigma-\ri\nu}{2},q)\, \vartheta_4(\frac{u-\ri \sigma-\ri\nu}{2},q)}\ ,\\
&&{\rm H}=\frac{1}{\sqrt{G_{uu}}}\ \frac{\vartheta'_1(0,q)}{2\ri\, \vartheta_2(\ri\nu,q)\vartheta_1(u,q)}\ 
\bigg[\, \vartheta_2(u-\ri\nu,q)\, \re^{\ri (\phi_+-\phi_-)}-
 \vartheta_2(u+\ri\nu,q)\, \re^{-\ri (\phi_+-\phi_-)}\, \bigg]\  .    \nonumber
\eea
Note that $\Omega^{+-}-\Omega^{-+}=\frac{1}{2\sqrt{G_{uu}}}\ \partial_u\log(S)$,
where 
\bea\label{isissu}
S={\textstyle \frac{1}{2\ri }}\ ( e_v^{-} e_w^{+}-e_v^{+} e_w^{-})\ .
\eea
Therefore  the last equation in \eqref{xosososa} and
\eqref{osososp} yield
\bea\label{ikoisoa}
\partial_u\log(S)=
\frac{\vartheta'_1(0,q)}{2 \vartheta_2(\ri\nu,q)\vartheta_1(u,q)}\ 
\bigg[\, \vartheta_2(u-\ri\nu,q)\, \re^{\ri (\phi_+-\phi_-)}+ 
 \vartheta_2(u+\ri\nu,q)\, \re^{-\ri (\phi_+-\phi_-)}\, \bigg]\ ,
\eea
or, equivalently,
\bea\label{aisisa}
S=\frac{1}{g^2}\ \rho(-u)\rho(u)\ \ \frac{\ri\, \vartheta_2(\ri\eta,q)\,  \vartheta_1(u,q)}
{ \vartheta_1(\ri\eta,q)\, \vartheta_2(0,q)}\ .
\eea
Here we use the function $\rho(u)$ defined in \eqref{kksisisa} and
$g$ is some $u$-independent  real constant.
Let us introduce  $E_\mu^\pm$ such that
\bea\label{ksosisa}
e_\mu^{+}= g^{-1}\  \re^{- \ri  \phi_+ }\ \rho(u)\ 
 E_\mu^+\, ,\ \ \ \ \ \ \ \ \ \ \ \ \ \ \ \ \ 
e_\mu^{-}= g^{-1}\  \re^{ \ri \phi_+ }\ \rho(-u)\ 
 E_\mu^-\ .
\eea
They satisfy the conditions
\bea\label{soiposa}
E^{-}_v E^+_w- E^{+}_v E^-_w=-2\  \frac{\vartheta_2(\ri\eta,q)\, \vartheta_1(u,q)}
{\vartheta_1(\ri \eta,q)\, \vartheta_2(0,q)}\ ,\ \ 
\ \ \ \ \ \ \ \ \ E^{-}_\mu=( E^{+}_\mu)^*
\ ,
\eea
and also
solve  a  system of differential equations
\bea\label{sisisaiusa}
&&{\rm W}\big[\,E^\pm_w,\,E^\pm_v\big]=\mp \ \frac{\vartheta_2(u\pm \ri\eta,q)\, \vartheta'_1(0,q)}
{\vartheta_1(\ri \eta,q)\,\vartheta_2(0,q)}\ \\ 
&&{\rm W}\big[\,E^+_w,\,E^-_v\big]+{\rm W}\big[\,E^-_w,\,E^+_v\big]=
2\ \frac{\vartheta'_2(\ri\eta,q)\, \vartheta_1(u,q)}{\vartheta_1(\ri \eta,q)\,\vartheta_2(0,q)}\ ,\nonumber
\eea
where ${\rm W}$ stands for the Wronskian, ${\rm W}[F,G]:=F\partial_u G-G\partial_u F$.
The system of 
Eqs.\eqref{soiposa},\,\eqref{sisisaiusa} can be integrated explicitly,  yielding the following expressions for $e^\pm_\mu$:
\bea\label{issasaus}
e^\pm_v&=&g^{-1}\ 
\re^{\pm \ri(\phi_+-\frac{\pi}{2})}\  \rho(\pm u)\ 
\bigg(\, a\  \frac{{\vartheta}_{4}(\ri\eta\pm u, q^2)}{{\vartheta}_{4}(\ri\eta, q^2)}-
b\  \frac{{\vartheta}_{1}(\ri\eta\pm u, q^2)}
{{\vartheta}_{1}(\ri\eta, q^2)}\,\bigg)
\nonumber\\
e^\pm_w&=&g^{-1}\ 
\re^{\pm \ri(\phi_+-\frac{\pi}{2})}\  \rho(\pm u)\  
\bigg(\, c\  \frac{{\vartheta}_{4}(\ri\eta\pm u, q^2)}{{\vartheta}_{4}(\ri\eta,q^2)}-
d\  \frac{{\vartheta}_{1}(\ri\eta\pm u, q^2)}
{{\vartheta}_{1}(\ri\eta, q^2)}\,\bigg)\ ,
\eea
where real integration constant $a,\,b,\,c$ and $d$ obey a single constraint $ad-bc=1$.
Using   $SL(2,R)$ coordinate
transformations \eqref{osososa}, 
we can bring the solution to the   forms  with either  $a=d=1, \ b=c=0$, 
or  $a=d=0, \ b=-c=1$. Since this two cases related by the coordinate transformation $(u,v,w)\leftrightarrow (-u,w,v)$, we  
accept the form
\eqref{uissasaus} without loss of generality.
Note that in Eq.\eqref{uissasaus}  we use the  constant  $l$ which 
substitute  the metric coefficient $G_{uu}\ \eqref{sososasq}$:  $\sqrt{G_{uu}}= \frac{l}{g}>0$. 
The ambiguity in sign of $e^3_u$ can   be  resolved by means of the condition
$\sqrt{G}:=\det(e^a_\mu)\geq 0$ that picks up  an orientation for  the vielbein.

\end{document}